# A lightweight cryptography (LWC) framework to secure memory heap in Internet of Things

Mahmoud Khalifa [a], Fahad Algarni [a,*], Mohammad Ayoub Khan [a], Azmat Ullah [b], Khalid Aloufi [c]

[a] *College of Computing and Information Technology, the University of Bisha, Saudi Arabia*
[b] *Department of Computer Science, La Trobe University, Australia*
[c] *College of Computer Science and Engineering, Taibah University, Saudi Arabia*

**Abstract** The extensive networking of devices and the large amount of data generated from the Internet of Things (IoT) has brought security issues to the attention of the researcher. Java is the most common platform for embedded applications such as IoT, Wireless Sensors Networks (WSN), Near Field Communications (NFC) and Radio Frequency Identification (RFID). The object programming languages such as Java, SWIFT, PHP and C++ use garbage collection after any object run which creates security loophole for attacks such as Next Memory Address Occupation (NMAO), memory replay, Learning Tasks Behaviors (LTB). The security risk increases in IoT when attacks exceeds the target device to the surrounding connected devices. Inappropriate or wrong operations causes energy loss and increased costs. In this paper, a security method to protect IoT system operation from memory heap penetration and address modification attack is proposed. The proposed method prevents directed attack by encrypting the object Garbage Collection at run time. To form a unique signature mechanism, the Cryptographic Hash Function (CHF) which employs a specific one-way hash algorithm. The proposed framework uses L-function based ECC and one-time Key (OTK) to secure the memory heap. Our method is used with open system where the effect on the operating system is not considered. The proposed method proved to be powerful and efficient which can help in achieving higher levels of security across several IoT applications, by enabling better detection of malicious attacks.

## 1. Introduction

Internet of Things (IoT) devices perform the operations and commands that the user selects remotely. Often IoT hardware is connected to devices with limited functions and simple installation. This is due to the limited used functions by users.

---

* Corresponding author.
E-mail address: fahad.alqarni@ub.edu.sa (F. Algarni).
Peer review under responsibility of Faculty of Engineering, Alexandria University.

Accessing IoT devices remotely is achieved through manual operations that assumed to be predefined in the system. After login to the IoT application interface, the user selects a particular device from the network and selects one of the installed tasks to activate or deactivate the device based on the desired service. The problem lies in the failure to choose the device or special task while accessing the IoT devices' functions. The aim of this paper is to discuss the complicated challenge of object location changes in physical address and the garbage collection method that erase any object after the binding in run time, and physical addresses changes that can be learned or achieved by a third party after session attack. Every programming language has its way to manage objects in memory heap as well as how to develop their tools to overcome bugs and vulnerabilities. Although software analytical techniques can detect several such flaws, a differential test uses similar programs to identify semantic bugs that do not exhibit specifically incorrect behaviors, such as statements or crashes. [1]. Previous attempts by scholars have addressed these issues by proposing NEZHA that is an effective framework for evaluation [1]. Another study addressed memory encryption problems for the architecture of the operating system to include code and data confidentiality with several vulnerabilities through the software stack [2]. Their solution uses complex hardware improvements that make it possible to carry out both encryption and decryption according to a trusted process. In the context of three general problem corrective methods, the existing memory encryption literature can also be reviewed. In addition, special kind of IoT attack has been discussed for on/off attacks of a malicious device which perform normal and abnormal services randomly to avoid being rated as a low trust node [3]. A Smart Middleware study was presented to assess the automatic trust of IoT resources and the attributes of service providers to avoid on–off attack.Table 1.

The security threats increase in cases when third party attempt to compromise session key utilizing brute force or guessing attack to violate corresponding heap address which point to the task name or object thread. Consequently, the content retrieved from memory location is relevant to the current task address, which will be then ready for next step of execution. In case of executing any unwanted task or stopping any other working ones, will also cause the waste of resource attack. Both time waste attack or service denial attack can cause negative impacts to services (i.e. forcing them to be repeated or break).

The risk of such threats can drastically increase when the executed tasks cause burning or damaging to the relevant devices. Any failure to effectively control the impacted equipment or services may lead to an extended disaster affecting the neighboring devices. Thus, in response to such threats, our contributions in this paper are: a proposal to solve the problem of how to overcome the Next Memory Address Occupation Attack (NMAPA) which normally enables intrusion to hack the running session and compromise NMAO. The compromised NMAO normally result in exploiting Garbage in the memory heap through Learning Tasks Behavior (LTB). The built model describes the way the system penetrated, where the hacker exploits the run time between the selection of the devices and the time of choosing a specific function. In addition, we propose a security model to the operation of the application from penetration or modification by using a simple garbage encryption method to be in the bottom of any object code and executed before the end of run time. This security model can enable better detection of malicious attacks targeting IoT enabled environments. Because of continuous IoT monitoring requirement, our proposal considered a reliable IoT attack prevention. It ensures that no heap address is sent to the enemy thread that is using guessing, brute force, or (NAOM) attack. It uses encryption for garbage collection at the end of any object run time. In the proposed method there is no way to find a message that produces a given hash even to attempt a brute force search attack of possible inputs to see if they produce a match or use a rainbow table of matched hashes. The one-time key (OTK) and cryptographic hash function (CHF) are not vulnerable to replay attacks. A brie classification of lightweight cryptography (LWC) is presented in Fig. 1.

The contribution of the work can be summarized as below:

1) Our work includes investigation of a broad spectrum of lightweight cryptography algorithms such as symmetric, asymmetric and one-time key (OTK) algorithms along with lightweight hash functions. Research gaps are identified on the resource constraints of an IoT system and solution has been proposed
2) A framework for OTK and L-function based ECC have been proposed to secure the garage collection. A simulation and performance analysis of the proposed algorithm with other ciphers has been carried out in terms of computational speed.

Table 1  Attack type and methods.

| Attack Type | Used method | |
|---|---|---|
| | TPM | LWC-ECC |
| Integrity | | |
| Denial of Service (Dos) | x | √ |
| Replay attack | x | √ |
| Parallel Session | x | √ |
| Platform Impression attack | | |
| Device Impersonate | √ | √ |
| Man in the middle | √ | √ |
| Run time attack | | |
| NMAO | x | √ |
| LTB | x | √ |

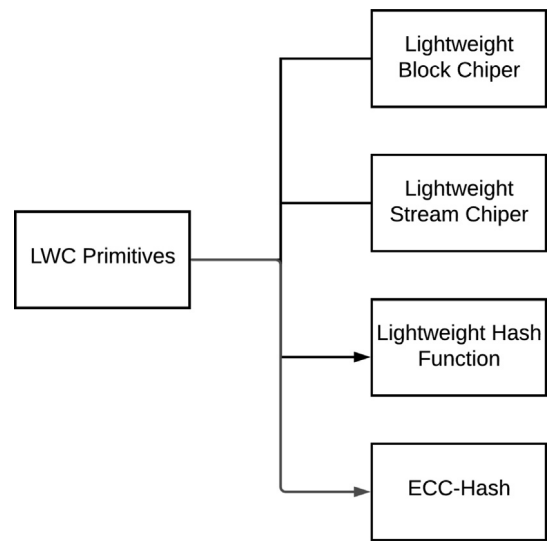

Fig. 1  LWC Primitive for Internet of Things.

The paper is organized as follows: Section 2 presents related work while the proposal of secure framework has been presented in Section 3. The simulation results and performance analysis have been discussed in Section 4. Finally, Section 5 concludes the paper along with future direction.

## 2. Related work

Recent research has shown that library language structure and module feature are susceptible to cyber-attack [4], where the form of confusion, sometimes combined with free-to-use, is the key attack vector that compromises modern C++ applications such as browsers or virtual machines. To specifically check all kind of controls, the authors propose to replace static controls with complete runtime. Further analysis analyzed the methodology for comparing a wide variety of uniform nomenclature control flow integrity mechanisms: first, qualitative discussion of conceptual protection guarantee, second, an assessment of quantitative safety of, third, an empirical assessment of its efficiency in the same test environment [5].

Other scholars have different evaluation mechanisms that employed the generation of equivalence classes and runtime performance of many machine learning techniques, which have been then evaluated to predict threats on the IoT [6]. The primary aim of this system is to build an IoT based infrastructure which is intelligent, stable and reliable, detects its vulnerabilities, secures a firewall against cyber-attacks and automatically recovers.

### 2.1. Attestation model

A number of remote software-based attestation processes, depending on strict time limits and other unfit to use in IoT systems, have been put in place. A dual attestation model, incorporating two phases, was proposed: static and dynamic [7]. The static attestation process tests the memory of the certified unit. The dynamic attestation technique checks the execution of the application code and can detect attacks during runtime. More recent attention has focused on proposing the configuration for coprocessor called Trusted Platform Module (TPM) which is extensively available in many hardware [8]. The module is owned by Trusted Computing Group (TCG). It helps in examining a certain system from a separated hardware and then enables the tamperproof memory locations to store system's sensitive information. However, this method was found to be incapable of measuring applications' behavior during runtime. This is extremely important in order to discover runtime malicious attacks including return-oriented programming (ROP), stack buffer overflow, and other relevant attacks.

A Multiple Tier Remote Attestation (MTRA) protocol has been designed [9] by the utilization of different types of networked IoT devices in terms of resources and computational capacity. More powerful TPM devices will be authenticated via trustworthy hardware and other devices will be checked using a software attestation. In addition, a randomized memory region for attestation is used with MTRA in order to increase the entropy of attestation responses. The attestation protocol could be viewed as a versatile way of verifying the integrity of a heterogenous IoT settings.

### 2.2. Authentication model

Recent research on authentication has employed two phases of security stage to encrypt password saved in external device, in order to enable it to be plugged and played at real program run time, with the aim of avoiding transmitting plain text passwords over the network [10]. This mode of operation maintains secure connection between user and device. Improved authentication model of users' identity based on Kerberos environment was used. Moreover, external memory was used to ensure external connection attack prevention and the TPM to interact safely with the internal components.

### 2.3. Memory model and attacks

Java Virtual Machines (JVMs) utilizes just-in-time (JIT) compilers to generate Java bytecode to machine code. However, it is also vulnerable to security threats with growing use of virtual machines. Heap spray attack discussed in [11] reveals the attacker is replicating malicious code in the heap and exploits the vulnerability to leap to the malicious code in the heap, allowing the execution of arbitrary code. The scholars proposed a mechanism that enables better detection and prevention of several heap spray attacks using Rand Heap. In addition, the researchers have argued that malicious attackers can gain a path to interrupt the whole organization network due to a specific failure in the relevant network at a centralized point [12]. In order to discover weaknesses in a hardware switch, two experiments were performed to find out whether basic generation of fake and suspicious software switches within the network is sufficient to exaggerate the JVM heap to its maximum capacity and result in crashing the relevant controller. These experiments performed utilizing a special Software Defined Network (SDN) supported switches and servers that can handle huge amount of data. It aimed at testing the likelihood of exaggerating the JVM heap to its maximum volume and determined a complete failure for such an approach. However, the researchers have identified that through affecting the garbage collector to pause performance of young generation garbage collections and activate the performance of full garbage collections, the denial of service was accomplished. This shift to full garbage collections, was due to the malicious attack exaggerating the heap memory assigned to the relevant flood light controller. By executing the performance of full garbage collections, the garbage collector can prevent the heap from being exhausted in due courses. This will result in an increased in both duration and frequency of the full garbage collections while it has been entirely executed [13]. For each full garbage collection, the time devoted started to exceed a single second, while it takes thousands of seconds at regular procedure. The control flow integrity is another method to evaluate programmer's behavior, in pro-grammer dynamic analysis [14]. By detecting control flow integrity, the programmer's runtime data, information and execution status can be obtained accurately. The combination of control flow integrity and dynamic trusted computation will allow the reliable verifier to be presented as a dynamic TPM-based measurement tool for control flow integrity. Compared to conventional measurement methods, they operate with greater precision and reliability, where programmers can be completely implemented in predefined control flows, while software solutions remain exposed to several different attacks [15]. The system parts either been protected nor monitored. The researchers proposed a solution to examine the targeted system from a remote hardware and safe the sensitive information of the system in its specific memory locations to overcome the problem. One of the widely known static techniques is Integrity Measurement Architecture (IMA). It aims at generating attestation from the kernel towards the application layer. Nevertheless, this technique is not capable of evaluating the applications' runtime behaviors, which is essential to identify runtime malicious attacks, for instance stack buffer overflow and ROP. In addition, there was an extended effort made to enable attestation protocol to cover both the verification process and the dynamic behavior collection. Normally the dynamic behavior is authenticated through the employment of different machine learning algorithms. The results have justified the success employment of such a technique and demonstrated higher detection rate for data sets weaknesses in web browsers such as Firefox.

Notwithstanding what has been implemented, including technologies used for attack and prevention, such as program analysis, memory management and machinery learning, our paper aims to establish an efficient method for better protecting waste collection and preventing attacks in IoT environments.

### 2.4. Threats in Internet of Things

The most outstanding and most critical of today's cyber-attacks is the distributed denial of service. It was considered a powerful attack on the existing Internet community. A survey was presented on distributed denial of service attack, prevention and mitigation techniques. [16]. Secure on-demand access that uses near-field communication, where a new mechanism is presented to secure element-based shared authentication and IoT attestation with a user computer, such as a mobile device. A newly proposed mutual locality authentication and attestation protocol will end anonymous mutual authentication between secure elements and the associated remote confidence attestation [17].

Microcontrollers are commonly used in many devices and will only become more common as the deployment of IoT grows [18], whereas IoT devices will increasingly perform functions that generate contemporary security and safety risks with more useful devices. Modern security protection mechanisms, aimed at preventing the execution of injected shell code, have emerged as a more secure solution to malicious code after such attacks. The technique used to take over program execution by modifying a function's return address via an exploit vector. This is accompanied by returning to small segments of code placed one by one in an executable memory to execute the plans of the attacker called return-oriented programming (ROP). It shows that this technique could fully regulate the Tiva TM4C123GH6PM microcontroller using a Cortex M4F processor. Normally, an appropriate code is loaded on Tiva microcontrollers to erase and rewrite the flash memory where the intended program resides. Then an arbitrary execution would be enabled by searching the ROM for a complete gadget package. This could allow an intruder to re-use the microcontroller by modifying the original functionality into his planned malignant behavior [18].

The complex connectivity and heterogeneity of the cyber physical systems components together with the exposed nature and error prone of the involved devices and tough operational settings is the gate of escalating vulnerability to malicious security attacks. Various reliability threats appeared as temperature induced dark silicon problem, soft errors, and process variation that caused many challenges that led to the advancement of many reduction techniques on various layers of the cyber physical systems/IoT stack. Likewise, malicious threats such as the manipulation of hardware elements, communication mediums and relevant software can cause the establishment of many detection and protection procedures on various layers of the cyber physical systems/IoT components [19]. The employment of Machine Learning in CPSs and IoT systems are due to the increases of the complexity and functional requirements. Similarly, wireless sensor networks (WSNs) are subject to many malicious attacks, such as node capture attacks, where an attacker physically captures, reprograms and redeploys a node in the network in hostile environments for applications such as battlefield surveillance [20]. A different approach was proposed to verify program integrity protocol to detect if a node is captured. The cluster head is tested by comparing the sensor node program memory contents before and after capture. The TPM-enabled program integrity verification protocol uses the hash key and pseudorandom-based dynamically computed function for network sensing node. The security review in the program integrity check protocol shows that there is a marginal possibility of a catching node that eludes the program integrity check and leaks the confidentiality of every uncaptured node. The proposed program can discover the captured node even in the presence of a strong opponent able to escape additional memory in order to verify the program integrity, it shows improvements in program integrity testing performance. The protocol is also enhanced in terms of low communication, computation and overhead storage compared with the related protocols for WSN program integrity checks.

## 3. Proposed framework for security memory heap

The problem discussed in this work is how to overcome the Next Memory Address Occupation Attack (NMAPA), that is normally able to beat the running session and exploit Garbage in heap attack through the process of Learning Tasks Behavior (LTB). There are two steps to solve the problem. First, hardware authentication step where using TPM to perform a protection process from the designated user address and prevent hacking operations of impersonating outside party. Second step is to make access violation harder by using a light weight cryptographic hash function (CHF) for garbage collection in the running object as shown in Fig. 2. A brute

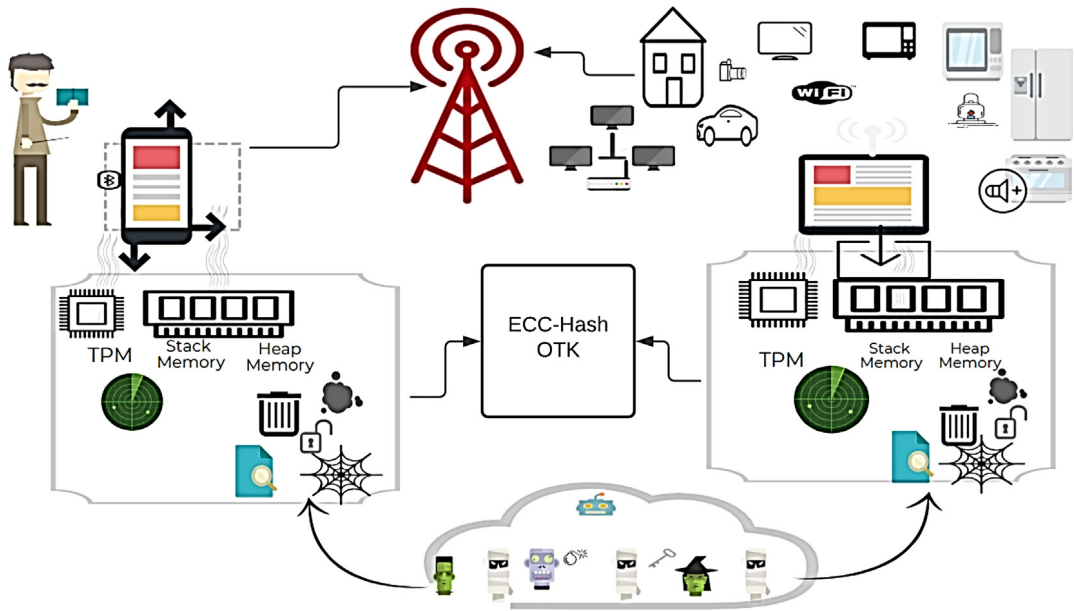

Fig. 2 LWC Framework to secure memory.

force attack or guessing by learning, are used to violate corresponding heap address.

The content retrieved from memory location is the task content or the object thread address. Firing any unwanted task or stopping any working thread, at least causes resource waste of time. Waste of time attack or service denial attack will cause services to be repeatedly on/off which cause system broke. The Worst-case if the executed task cause burning or damaging the IoT devices.

### 3.1. Garbage encryption

A stream-cipher algorithm based on one-time keys and robust chaotic maps was used to obtain high security and improve dynamic degradation [21]. Where the linear chaotic map is used as a pseudo-random keystream generator, where a color image encoding scheme has been introduced and the lightweight Hash encryption function is used as a one-way hash operation.

In this paper, we have presented the garbage collection encryption based on CHF that apply OTK to secure the memory. The proposed cryptographic method has higher security due to an extremely large key space generated from CHF.

### 3.2. ECC-Based hashing technique

Elliptic curve is one the popular and efficient lightweight technique that is based on the equation of points on elliptic curve as shown in Fig. 3. In this work we proposed CHF based on the L-function [22]. We can define has function as $H$ that takes variable length input and produces fixed length output to ensure the integrity as shown in equation(1).

$$H(.) : \{(x,y) \in E_p(a,b) \rightarrow \mathbb{Z} \quad (1)$$

The coordinates on elliptic curve is $E_p(a,b)$ and produce the integer value $Z$. The equation for elliptic curve can defined as follows [22] in equation (2) and (3):

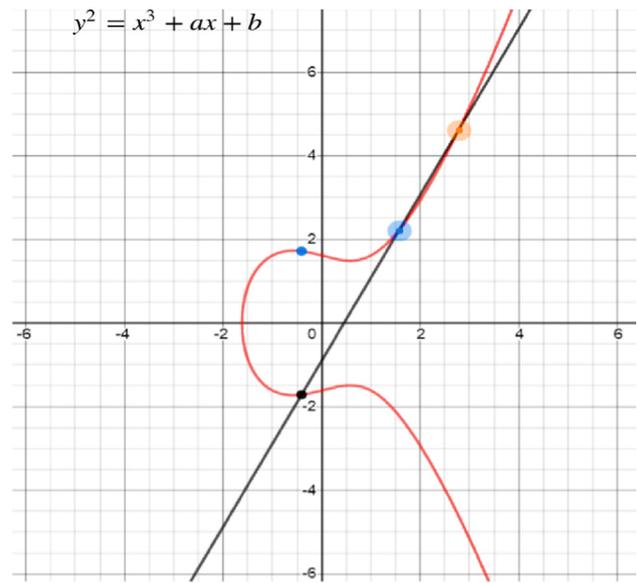

Fig. 3 Elliptic curve.

$$E_c : y^2 = x^3 + ax + b, (a,b) \in K \quad (2)$$

$$\Delta(E_c) = -4a^3 - 27b^2 \quad (3)$$

Now, we can apply Mordell-Weil theorem as follows show in equation (4):

$$E_c(K) = \{(x,y) \in \mathbb{K} \times \mathbb{K} | y^2 = x^3 + ax + b\} \cup O \quad (4)$$

$O$ is a pint at infinite as an identity element which is represented as abelian group with finite set of $\mathbb{K} = \mathbb{F}_q, q$ is a prime number such that $(a,b) \in \mathbb{Z}$. If we take prime number p, therefore, $a_p = p + 1 - \#E_c(\mathbb{F}_p)$. The $\mathbb{F}_p$ is has $p$ elements with following order as shown in equation (5).

$$E_c(\mathbb{F}_p) = O \cup \{(x,y) \in \mathbb{F}_p x \mathbb{F}_p | y^2 = x^3 + ax + b\} \quad (5)$$

Now, apply the Hasse theorem which is will shall satisfy $p, a_p$ for $|a_p| \leq 2\sqrt{p}$. After attaching L-function to $E_c$, we get following equation(6)[22].

$$L(s, E_c) \prod_{p | \Delta(E_c)} \frac{1}{1 - a_p p^{-s}} \prod_{p \nmid \Delta(E_c)} \frac{1}{1 - a_p p^{-s} + a_p p^{1-2s}} \quad (6)$$

The $p$ divided $\Delta(E_c)$, the $a_p = 0 \pm 1$. The singularity can be found using $E_c \bmod p$. Finally, the Euler product can be represented by the equation(7).

$$L(s, E_c) = \sum_{n=1}^{\infty} a_n n^{-s} \quad (7)$$

The set of alll elliptic curvers $E_c$ over $\mathbb{Q}$ with very large number(T) is represented by $\mathscr{L}^T$. The $\alpha, \beta$ be assumed as fixed postive numbers then $T \leq \Delta(E_c) \leq 2T$ should also hold such as $k$ and $m$ should satisfy below equation(8).

$$\log T^\alpha \leq k, m \leq (\log T)^\beta \quad (8)$$

Finaly, the CHF can be represented by equation(9) as follows:

$$H : \mathscr{L}^T \to \mathbb{Q}^{k+1} \quad (9)$$

Now, we can apply One-Time Key(OTK) algorithm to encrypt the garbage collection memmory. The OTK is an algorithm which uses a new key everytime a new encryption is performed. The generalized mathematical equation for OTK can be expressed as equation(10).

$$C_i = E(M_i, H_i \% m), i = 1, 2, 3, \cdots, n \quad (10)$$

Where $C_i$ is the cipher or encrypted text generated by the encryption operation. $E$ is the encryption operation, $M_i$ is the $i^{th}$ character of the plaintext, $H_i$ is the $i^{th}$ key genereated from the garbage collection. The $H_i$ is generated from equation(9). The $m$ is a modulo fixed number. In a similar way, we can also mathematically represent the decryption operation in equation(11).

$$M_i = D(C_i, H_i \% m), i = 1, 2, 3, \cdots, n \quad (11)$$

Where $D$ is the decryption operation, $M_i$ is the $i^{th}$ character of the plaintext, $H_i$ is the $i^{th}$ key genereated from the garbage collection.

## 4. Simulation and performance analysis

Encryption methods, such as AES, RSA and Elliptic Curve, run on distributed systems but require reasonable processing and capable memory capacities. However, these methods are not suitable for embedded systems and sensor networks [23]. One of its advantages is that it can use shorter keys for operation [24]. The cipher block used in [25] has been described as lightweight where the spectrum of lightweight ciphers and chip design space for low-resource devices has been accurately defined. This work explores the simulation of the LWC technique and contrasts it with selecting top-performance algorithms in different metrics, indicating that the LWC based technique is a good guide for software and hardware implementations to make the garbage memory more secure. The proposed framework uses the OTK and ECC for generating keys in IoT. The Fig. 4 shows the process how the physical address appears while the user selects IoT device and choose its task using a prototype model of home automation.

### 4.1. Security analysis

It is increasingly crucial that IoT is realistically secure from widely known vulnerabilities. Unless there are continuous improvements of the security posture for IoT devices, attackers will find ways to exploit vulnerabilities for their own desires. In case of a slipped virus in IoT system, everything done will caught by the hacker. The active session's data and its RAM heap location addresses will be sent out for learning how IoT devices works. Fig. 4. represents how users first select its device in their network then chooses the corresponding task. When the device memory heap session attacked, the address of the connected device in the relevant memory or the corresponding tasks changes. In addition, when a user moves to the device functions to choose a task, the services does not determine the address, or its tasks has been changed. A false selection causes a false selection of its function. Fig. 4. demonstrates the scenario of how IoT proposed system works. The new address will be either processed or delayed until the functional service has been selected by the attacker before the request is terminated and device does not show up or raise conflict error. In order to calculate the applications' behavior during runtime, which is required to detect several malicious attacks including ROP and stack buffer overflow in the proposed dual attestation mechanism of IoT systems software, to identify vulnerabilities during runtime. The authors have also developed attestation protocol with high-level-based attributes to cover both the verification process and dynamic behavior collection. Different machine learning algorithms are specifically employed to verify the dynamic behavior.

In this paper, we suppose that an attacker always tries to access the physical address of the running job. If session was created the attacker tries to read the session to compromise key using brute force or guessing attack. Attacker violate corresponding heap address which point to the task name or object thread. Content retrieved from memory location is the task, while being ready for execution step. Firing any unwanted task or stopping any other working one, at least causes waste resource attack. Time waste attack or service denial attack cause services to repeat or break. The Worst case when an executed task causes burning after damaging devices.

### 4.2. Prototype for the framework

IoT devices faces unique security challenges due to its constrained resources. Shortage of storage, manual processing handling dependence, limited memory and power of IoT objects do not support the deployment of advanced security protocols, which are often resource intensive, the use of Public Key technologies to ensure secure end to end communications between devices and services over the Internet [26]. We used OTKs to improve the efficiency of hash encryption protocols employed in IoT devices.

### 4.3. Security benchmark

It should not be possible for an adversary to detect two encrypted hash garbage collection with significantly similar numbers or to extract useful information on the data, given its hash only. We used Java NetBeans platform for the IoT prototyping and implementation of LWC.

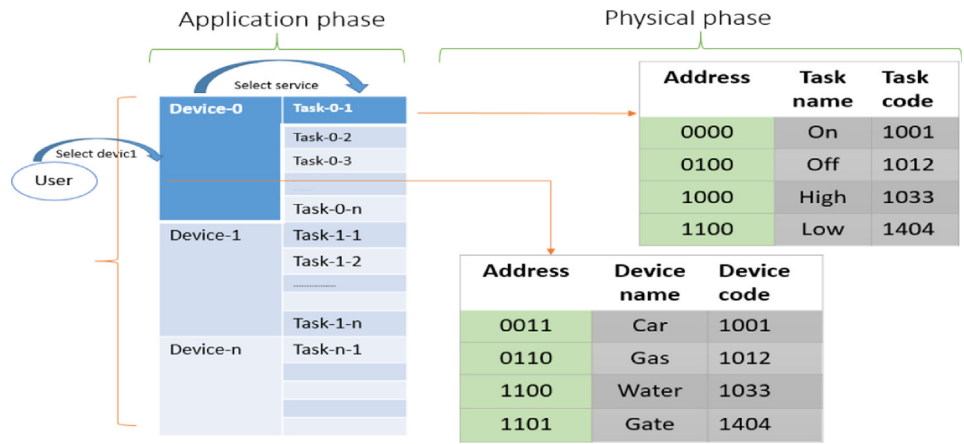

**Fig. 4** The procedure how the physical address appears while the user selects IoT device and choose its task.

Fig. 5. Illustrates the IoT environment and its devices that being selected and executed on/off where security was applied to prevent garbage collection attack. In the table we have presented a comparative analysis of features of TMP and LWC-ECC methods.

*Hash and OTK computation Time:* This indicate the time utilized by the ECC-based hash algorithm to create a hash from a garbage memory collection text and applying OTK. It is the difference between the end and start times of encryption and decryption as expressed in equation (12) and (13).

$$E_{(t)} = E_{end(t)} - E_{start(t)} \quad (12)$$

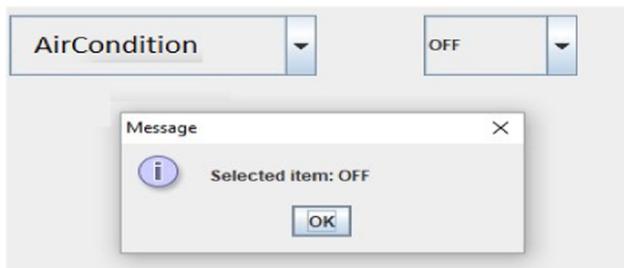

**Fig. 5** IoT devices selected and executed on/off.

$$D_{(t)} = D_{end(t)} - D_{start(t)} \quad (13)$$

where:
$E_{(t)}$ – Total hash time and OTK encryption time
$H_{end(t)}$ – Hash and OTK encryption end time
$H_{start(t)}$ – Hash and OTK encryption start time
$D_{(t)}$ – Total OTK decryption start time
$D_{end(t)}$ – The OTK decryption end time
$D_{start(t)}$ – The OTK decryption starting time

Table 2 present comparative analysis of encryption and decryption time for different size of algorithm. We have analyzed computational time for AES with SHA-1 algorithm, RSA with SHA-1 and finally the proposed ECC-based-Hash and OTK.

From the Fig. 6, we can observe that average computational time for encryption process which are 1.0508 μs, 1.926 μs, and 2.187 μs respectively. The proposed ECC-based hash and OTK has minimum encryption time as 1.0508 μs. Similarly, we can observe that average computational time for decryption process which are 1.398 μs, 1.713 μs and 2.102 μs respectively. The proposed ECC-based hash and OTK has minimum decryption time as 1.398 μs as compared to other method.

## 5. Conclusion and future work

Any memory heap running application specified and controlled by the operating systems. Although there is embedded protection, there are still vulnerabilities. In this paper, we have presented a comprehensive and systematic method of how to overcome garbage collection vulnerabilities, where object runtime malicious attacks normally take place. We employed well known prevention techniques based on garbage collection vulnerability on how physical address in memory heap can be remotely read or learned. Our method was accomplished in two stages. In the first stage, we used TPM unit device to ensure internal authentication with processor from any unauthorized party. In the second stage, we developed and evaluated a cryptographic hash function (CHF) that encrypt any garbage collection object. The CHF utilizes OTK algorithm that effectively encrypts garbage collection, to prevent NMAPA, which is one of today's cyber world's most risky attacks. It poses an overwhelming challenge to the existing Internet culture with basic but extremely powerful attack technique. We have defined the main characteristics of memory heap attacks as well as the garbage collection encryption model, which enhanced security and stopped external and malicious attacks.

**Table 2** Computational time different algorithms.

| Heap Memory Size (KB) | Proposed ECC-Hash + OTK (μs) | | AES + SHA-1(μs) | | RSA-SHA-1(μs) | |
|---|---|---|---|---|---|---|
| | $E_t$ | $D_t$ | $E_t$ | $D_t$ | $E_t$ | $D_t$ |
| 128 | 0.512 | 0.423 | 0.923 | 0.823 | 1.235 | 1.122 |
| 512 | 0.591 | 0.491 | 1.213 | 1.113 | 1.413 | 1.343 |
| 1024 | 0.912 | 0.812 | 1.321 | 1.211 | 1.523 | 1.435 |
| 2048 | 1.512 | 1.401 | 1.823 | 1.712 | 1.993 | 1.862 |
| 4096 | 1.921 | 1.812 | 2.512 | 2.356 | 2.701 | 2.623 |
| 8192 | 2.153 | 2.014 | 2.563 | 2.423 | 2.934 | 2.878 |
| 16,384 | 2.952 | 2.832 | 3.125 | 2.935 | 3.512 | 3.453 |
| Average Time | 1.508 | 1.398 | 1.926 | 1.713 | 2.187 | 2.102 |

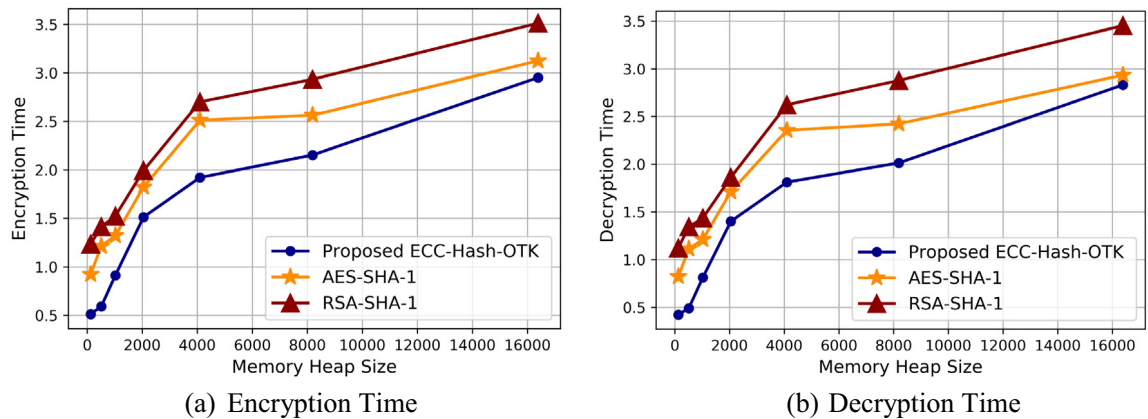

(a) Encryption Time      (b) Decryption Time

**Fig. 6** Comparative Analysis of computational time of different algorithms.

Future research could focus on developing similar models to detect and prevent emerging malicious attacks, considering the increasing demand of deep and integral solution between the operating system and the relevant applications.

**Declaration of Competing Interest**

The authors declare that they have no known competing financial interests or personal relationships that could have appeared to influence the work reported in this paper.